\documentclass[conference]{IEEEtran}
\IEEEoverridecommandlockouts
\usepackage{cite}
\usepackage{amsmath,amssymb,amsfonts}
\usepackage{algorithmic}
\usepackage{graphicx}
\usepackage{textcomp}
\usepackage{xcolor}
\usepackage{xspace}
\usepackage{subfigure}
\usepackage{textcomp} 
\usepackage{blindtext}
\usepackage{hyperref} 
\usepackage{balance}

\makeatletter
\def\ps@IEEEtitlepagestyle{%
  \def\@oddfoot{\mycopyrightnotice}%
  \def\@evenfoot{}%
}
\def\mycopyrightnotice{%
  {\footnotesize 978-1-7348995-2-8/22/\$31.00 ©2022 Immersive Learning Research Network\hfill}% <--- Change here
  \gdef\mycopyrightnotice{}% just in case
}
\def\BibTeX{{\rm B\kern-.05em{\sc i\kern-.025em b}\kern-.08em
    T\kern-.1667em\lower.7ex\hbox{E}\kern-.125emX}}

\makeatother    
    
\begin{document}

\title{An Explore of Virtual Reality for Awareness of the Climate Change Crisis: A Simulation of Sea Level Rise\\
}

\author{\IEEEauthorblockN{Zixiang Xu}
\IEEEauthorblockA{\textit{School of Computer Science, University College Dublin} \\
Dublin, Ireland \\
zixiang.xu@ucdconnect.ie}
\and
\IEEEauthorblockN{Yuan Liang}
\IEEEauthorblockA{\textit{School of Computer Science, University College Dublin} \\
Dublin, Ireland \\
yuan.liang@ucdconnect.ie}
\and
\IEEEauthorblockN{Abraham G.~Campbell}
\IEEEauthorblockA{\textit{School of Computer Science, University College Dublin} \\
Dublin, Ireland \\
abey.campbell@ucd.ie }
\and
\IEEEauthorblockN{Soumyabrata Dev}
\IEEEauthorblockA{\textit{School of Computer Science, University College Dublin} \\
%\textit{ADAPT SFI Research Centre, Dublin, Ireland}\\
Dublin, Ireland \\
soumyabrata.dev@ucd.ie }
% \and
% \and
}

\maketitle

\begin{abstract}
Virtual Reality (VR) technology has been shown to achieve remarkable results in multiple fields. Due to the nature of the immersive medium of Virtual Reality it logically follows that it can be used as a high-quality educational tool as it offers potentially a higher bandwidth than other mediums such as text, pictures and videos. This short paper illustrates the development of a climate change educational awareness application for virtual reality to simulate virtual scenes of local scenery and sea level rising until 2100 using prediction data. The paper also reports on the current in progress work of porting the system to Augmented Reality (AR) and future work to evaluate the system.
\end{abstract}


\section{Introduction}
Climate change is accelerating, bringing the world ‘dangerously close’ to irreversible change~\cite{fountain_2019} and even potentially affected the psychological wellbeing~\cite{CLAYTON2020102263} of people around the world. However, large portions of the public still believe that global warming is not happening~\cite{leiserowitz2020climate}, and only a minority believe it will impact them and their families severely over the next fifteen years~\cite{banks_2020}. This has much to do with the fact that the effects of climate change are "gradual" rather than "immediate", which may make people feel undecided or uncertain~\cite{Schuldt2016}. 
A substantial segment of the public ignore the urgency of climate change because of its inconspicuousness. This paper details a unique case study that has  targeted a  population where within 50 years their homes will be under water. This is within their or their children's lifetimes so has a direct effect on them but still many are skeptical.  It is hoped that through visualization in a VR application  which  dynamically simulates the sea level rise, finally it will become apparent how serious the problem is. For the families within this target group, this education will aid them in making the difficult but necessary decisions for their future and their children future.    

Compared with traditional media and educational media, Virtual Reality (VR), as an emerging science and technology, has an incomparable capability of information transmission due to its immersive nature. In 2018, Markowitz,D. M. \textit{et al.}~\cite{Markowitz2018} tested the efficacy of immersive VR as an education medium for teaching the consequences of climate change (Fig.1), and got positive results that displayed more positive attitudes toward the environment after comparing pre- and post-test assessments.

\begin{figure}[htb]
\centering 
\includegraphics[width=0.49\textwidth]{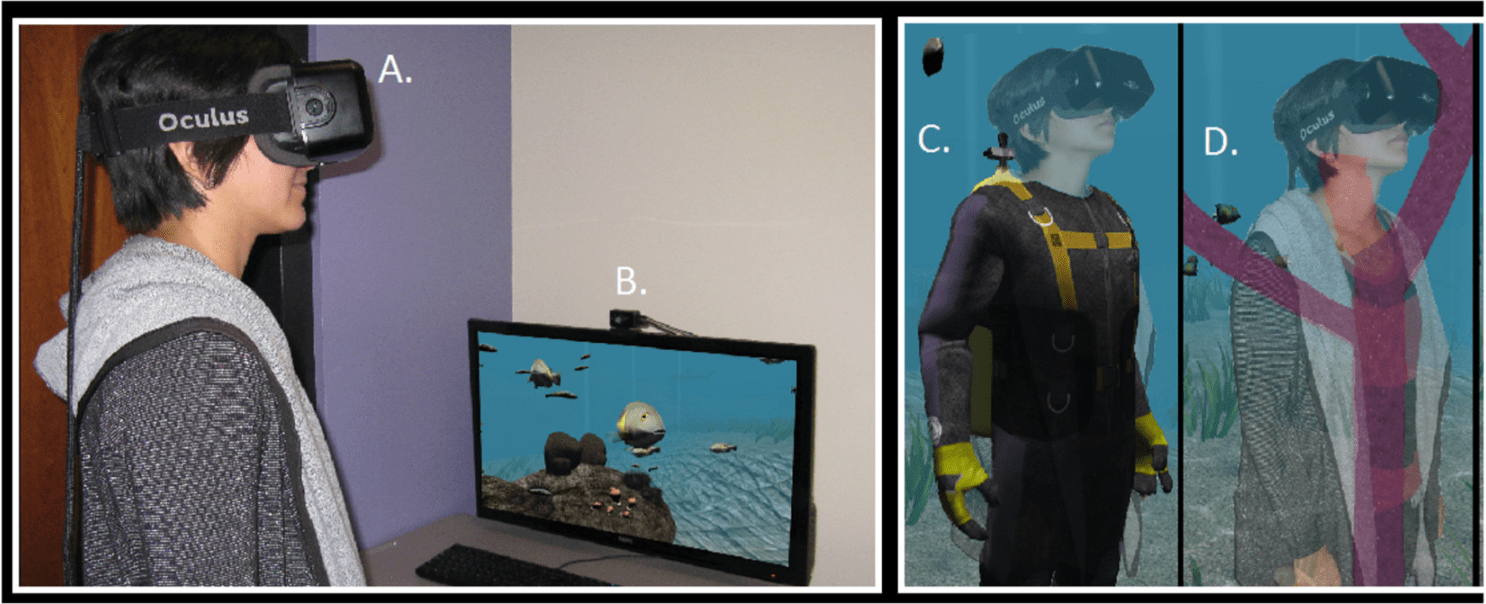} 
\caption{Markowitz,D. M. \textit{et al.} ~\cite{Markowitz2018} are testing their project.} 
\label{fig:fig7} 
\end{figure}

While Markowitz \textit{et al.} tried to simulate an immersive underwater world to show the process and effects of rising sea water acidity, this work is targeted at a much more local level. Specifically a local residential area, Portrane (a seaside settlement located in North County Dublin, Ireland).  The aim of this work is to give a Virtual Reality experience of sea level rising along the coast within the residential environment.  This work mirrors work in VR \cite{calil2021using} that has been  used to highlight sea level rise in another coastal communities such as Santa Cruz,California. 

Section 2 gives the development and implementation of the application. Section 3 will discuss the experiments so far that have taken place  and the plan of further ones. It also highlights future work including a demonstration of an Augmented Reality application that was ported from the VR project. Conclusion will be given in section 4.

\section{VR Application Development and Implementation}

\subsection{Application design with Unity3D and Oculus Quest}
 360°  immersive video applications offer the potential to provide vivid experiences especially in Cultural Heritage education which has been proven in past studies~\cite{Argyriou2020},\cite{vskola2020virtual} .This project origins come from a 360-degree fixed scene that is combined with a 360 video and a simulated 3D ocean model (Fig.2). This initial demo was developed to illustrate the simple sea level changing at a fixed view point and it can run on both mobile and Head-Mounted Device (HMD) like Oculus Quest and provide elementary immersive scene. But the fixed point limitation could not provide an immersive experience and a large number of panoramic video shooting and complex simulation of sea level change would have been required for every new viewpoint, and this tech is less responsive, less interactive than virtual reality~\cite{conroy_2017}, Therefore, we chose a VR application with virtual scenes as a better solution.

\begin{figure}[htb]
\centering 
\includegraphics[width=0.49\textwidth]{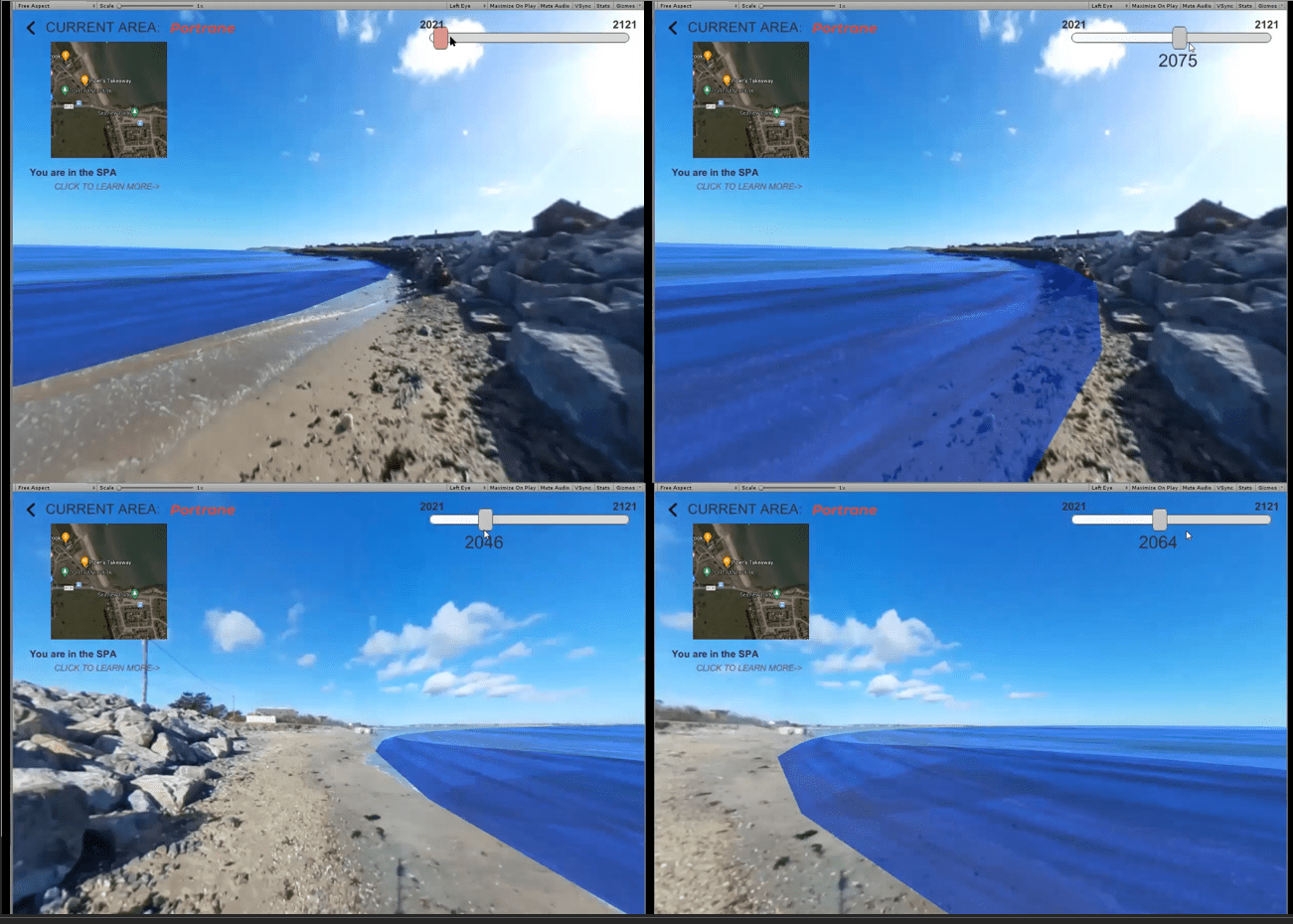} 
\caption{The 360-degree scene viewer based on punctuation video and ocean model.} 
\label{fig:fig1} 
\end{figure}

This application is developed with Unity3D game engine~\cite{Unity3D} and can run on the Oculus Quest platform~\cite{Oculus} (Fig.3). Oculus Quest 1 is the device used for development and it provides 1440 × 1600 resolution and the max offers a maximum refresh rate of 72fps. As a VR all-in-one headset machine, it doesn't need external computing or rendering support, but uses a Qualcomm Snapdragon 835 SoC with 4 GB of RAM (Three of the four 2.3 GHz CPU cores of the chip are reserved for software) and a Adreno 540 GPU to compute and render by its own, this is close to the graphics power of mid-low end mobile phones on the market. The controllers of Oculus Quest can provide all the interaction requirements in this application.

Five points of interests in the beach area of Portrane were picked to be included, and based on each a virtual scene with residential views was built. These points were chosen due to their previously identification as points of scenic beauty in the area. The home menu page provides a map with these 5 points of interests (Fig.4 (b)). The user interface interaction is based on the virtual laser that starts from the right controller and achieve the operations through the interactive buttons on the user interface pointed by the laser, in actual operation, users can achieve interaction by push button "A" on right controller when the laser hit the button on the interface. Users can pick any one of 5 point of interests to explore the virtual scenes, view the ocean level in specific year and learn about the corresponding ecological information (Fig.4 (e)\&(f)) for each different point of interests. The information shows as text, audio and video, contains special local ecosystem, local characteristics, species, the impact of climate change and how to make environmental protection. The controlling of ocean level rise simulation is designed as a slider for a easy and dynamic show of the sea level height. Specifically, the slider bar is designed to be divided into 80 consecutive points, each of which represents a year and pass a value for which a simulation model of the ocean will take in and rise to its corresponding position, users can also directly jump to 2021, 2050 or 2100 using the buttons below the slider bar (Fig.4 (c)\&(d)). In addition, a rotating compass is also added to the upper right corner of the user interface based on the geographical direction of the real world to indicate the direction in the virtual environment, helping users to better connect the virtual scene with the information in the real scene. Immersion in consideration, the lower left corner of the user interface adds a "hide UI" button, the user can at any time through the button to hide in addition to the button all the interactive interface, and click again to display, in order to in a selected years of rising sea levels scenarios to achieve better immersive experience.

\begin{figure}[htb]
\centering 
\includegraphics[width=0.49\textwidth]{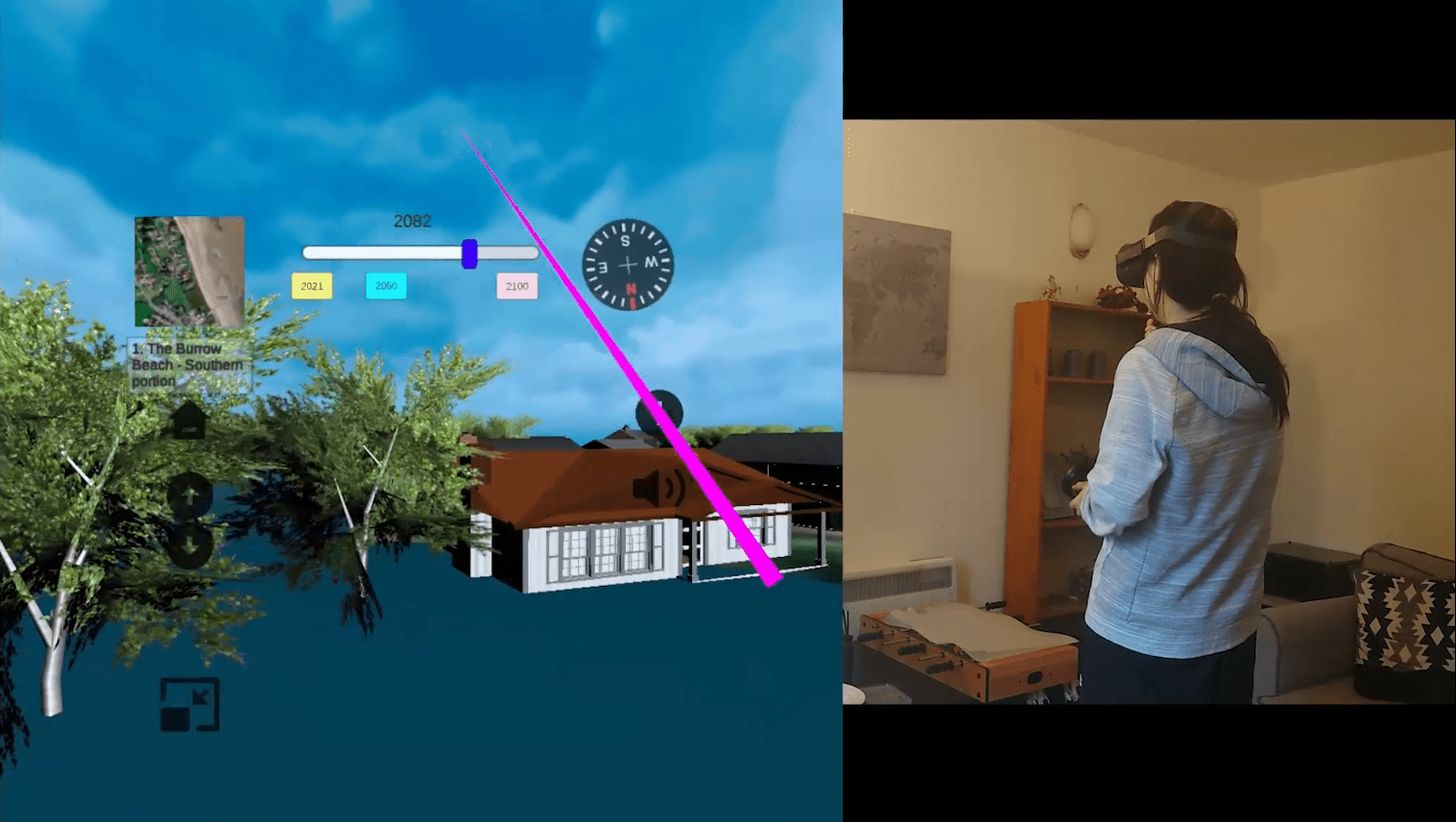} 
\caption{Using Oculus Quest Head-Mounted Device (HMD) to run the application and make interaction with user interface.} 
\label{fig:fig8} 
\end{figure}

\begin{figure}[htbp]
\centering
\subfigure[ ]{
\includegraphics[width=3.75cm]{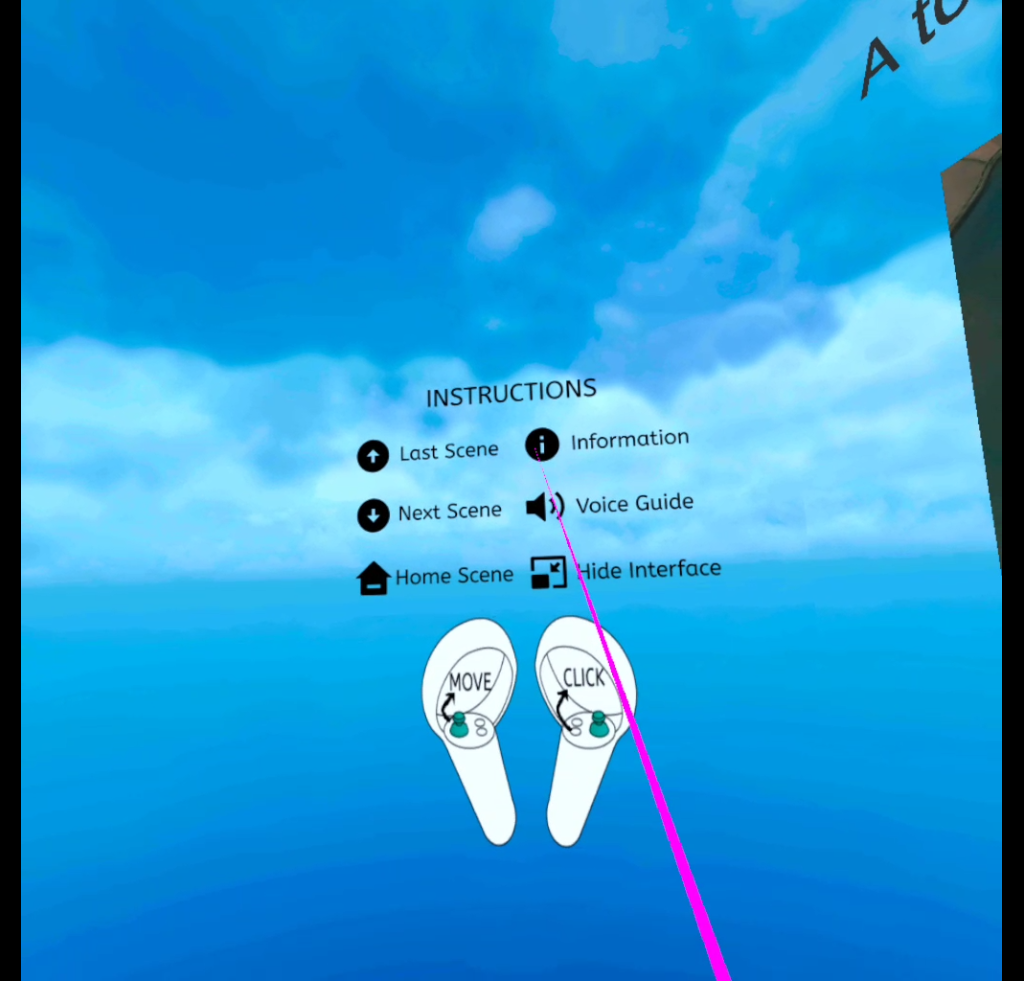}
%\caption{fig1}
}
\quad
\subfigure[ ]{
\includegraphics[width=3.75cm]{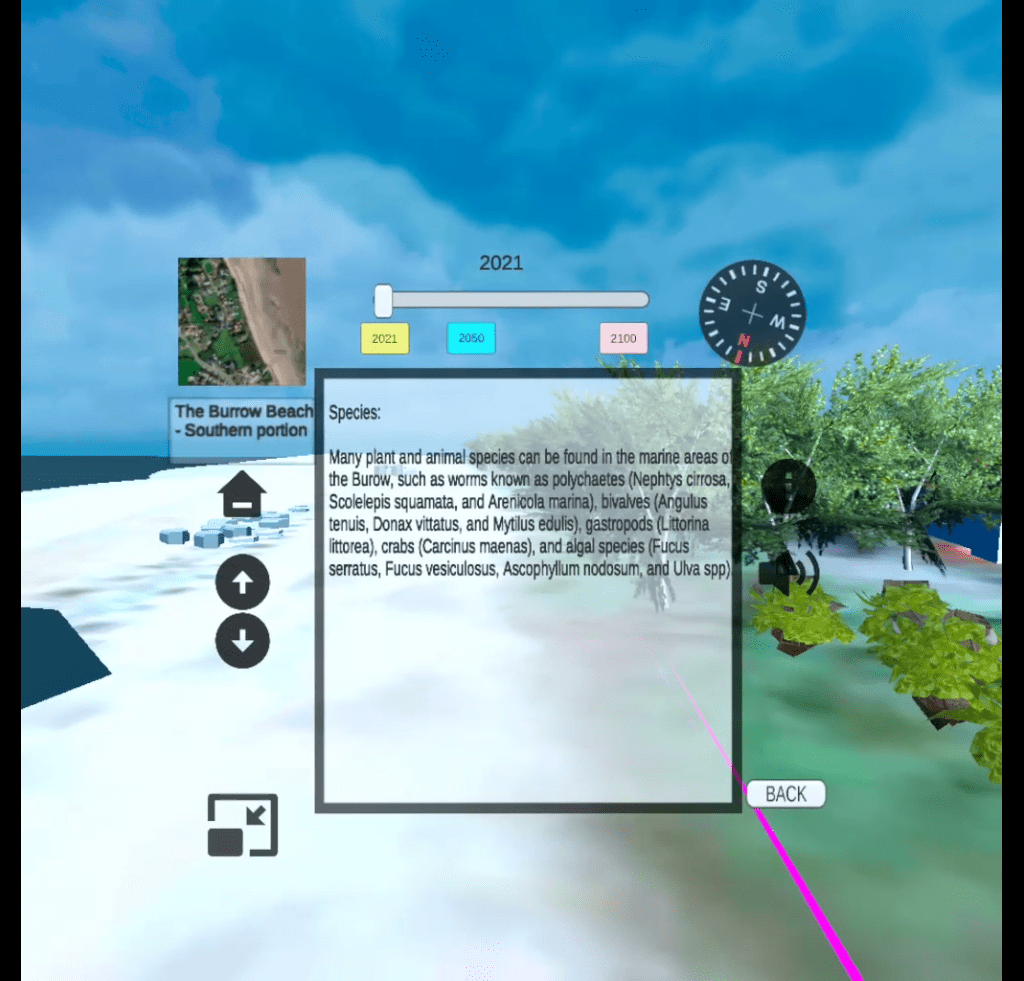}
}
\quad
\subfigure[ ]{
\includegraphics[width=3.75cm]{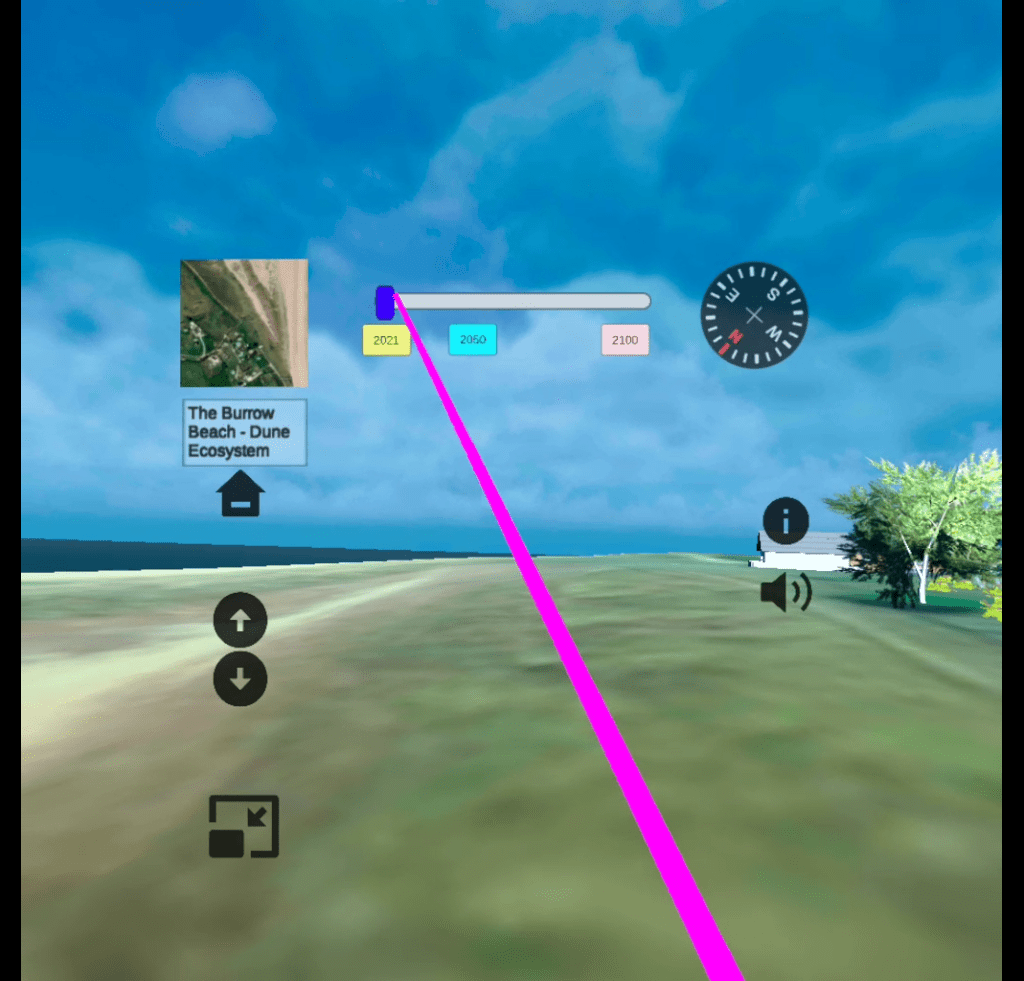}
}
\quad
\subfigure[ ]{
\includegraphics[width=3.75cm]{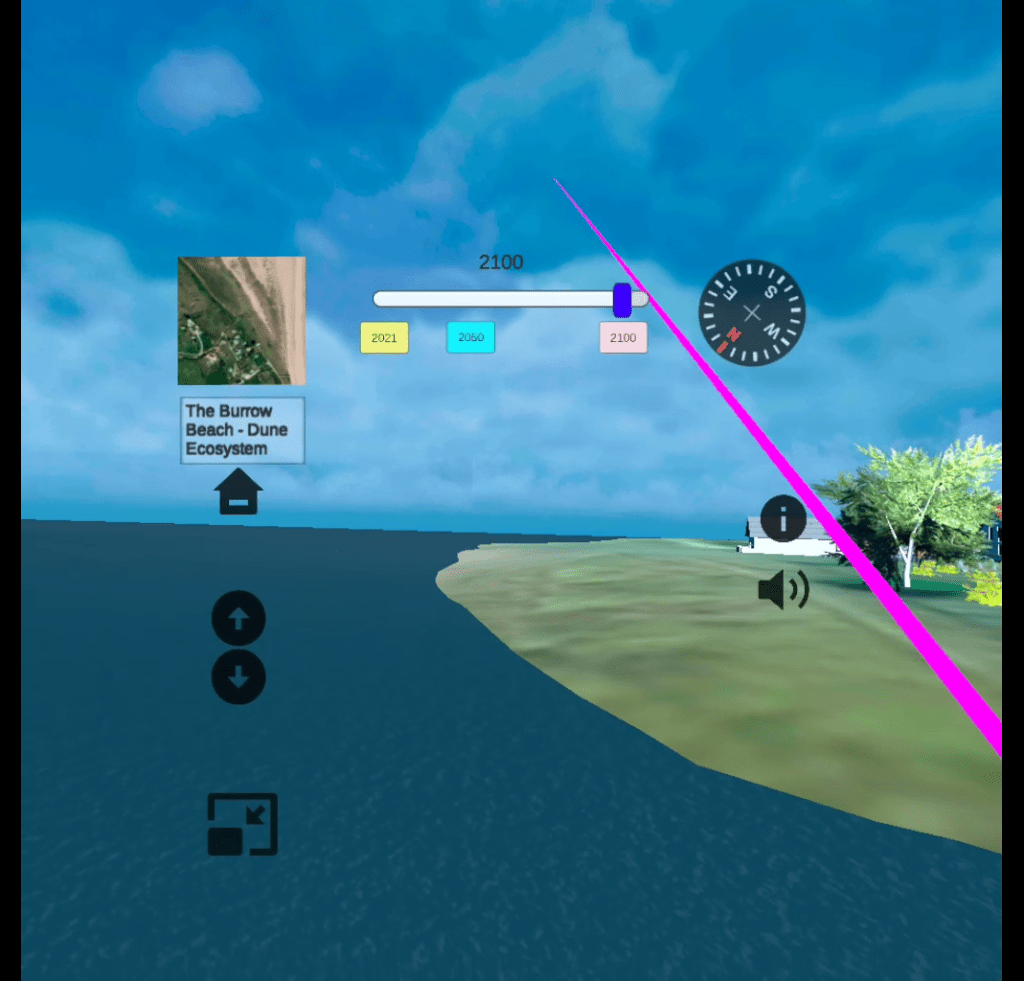}
}
\quad
\subfigure[ ]{
\includegraphics[width=3.75cm]{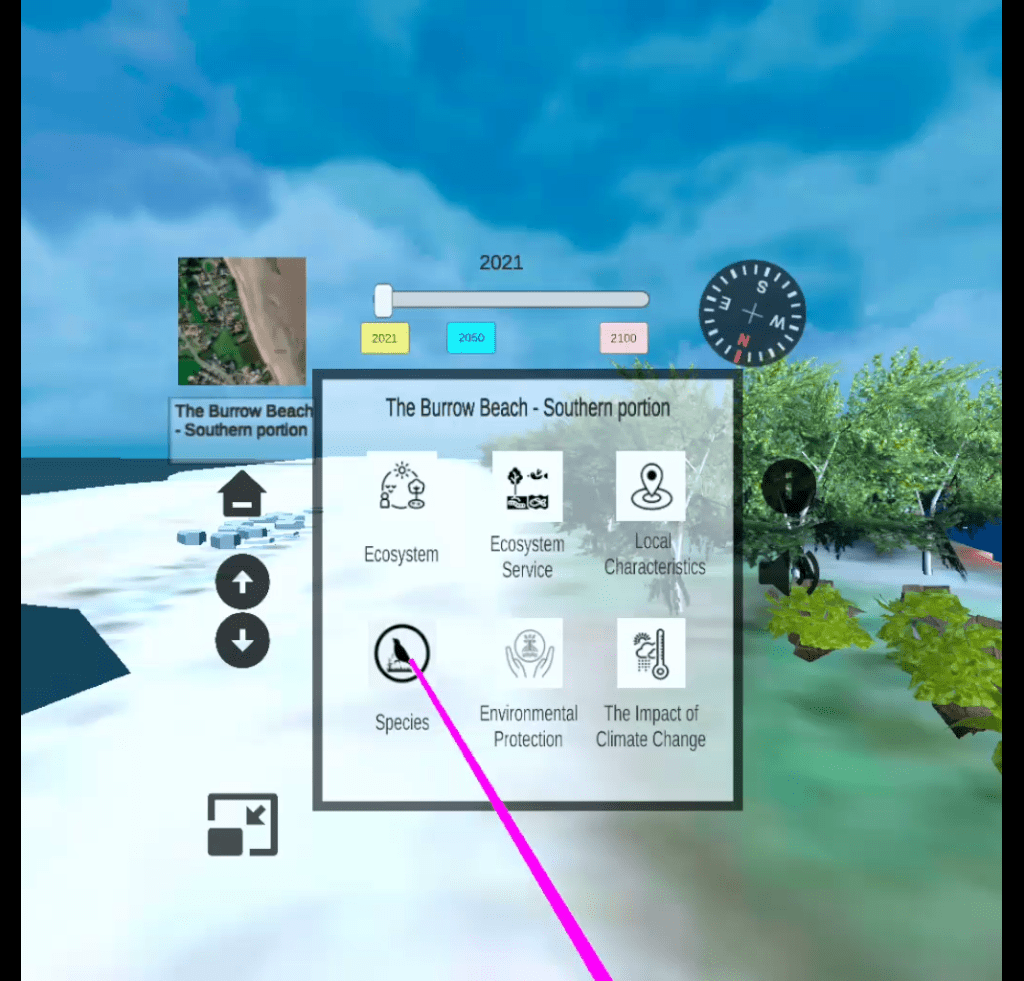}
}
\quad
\subfigure[ ]{
\includegraphics[width=3.75cm]{PosterPicture007.png}
}
\quad
\subfigure[ ]{
\includegraphics[width=3.75cm]{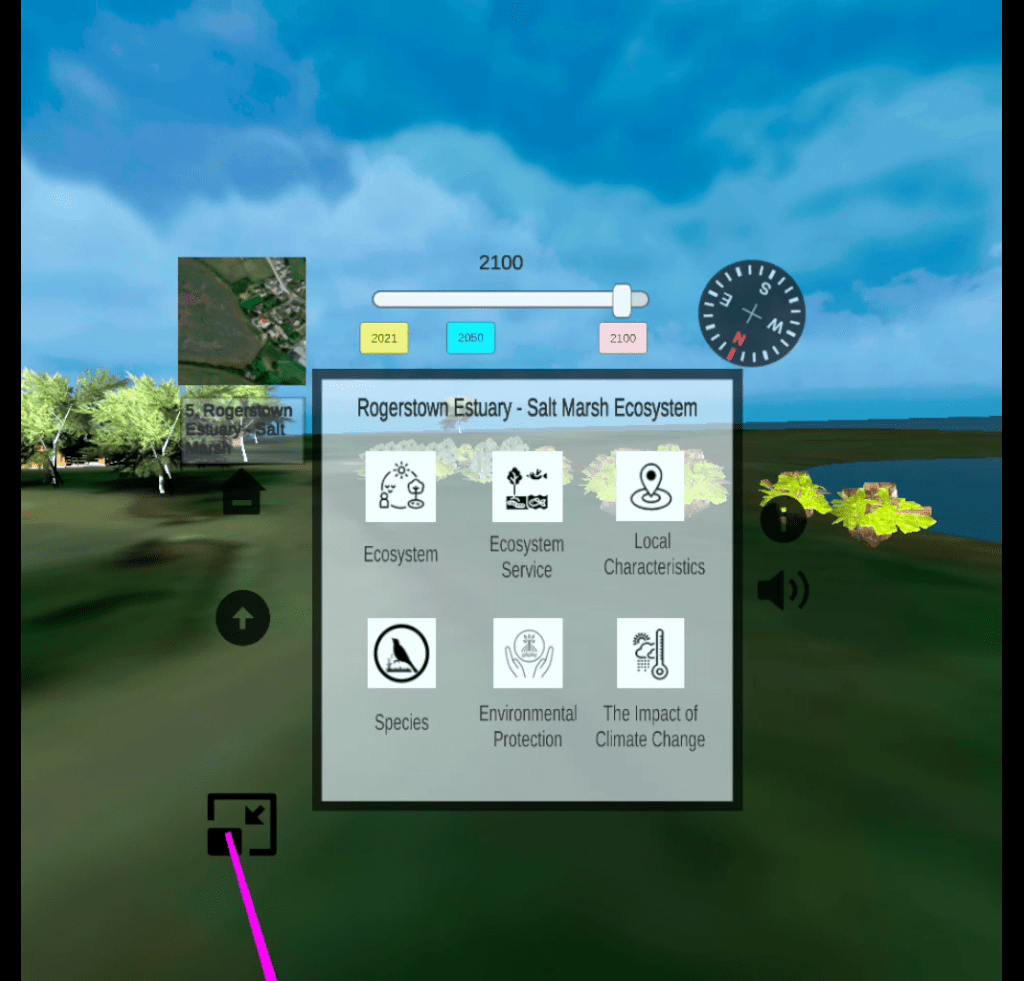}
}
\quad
\subfigure[ ]{
\includegraphics[width=3.75cm]{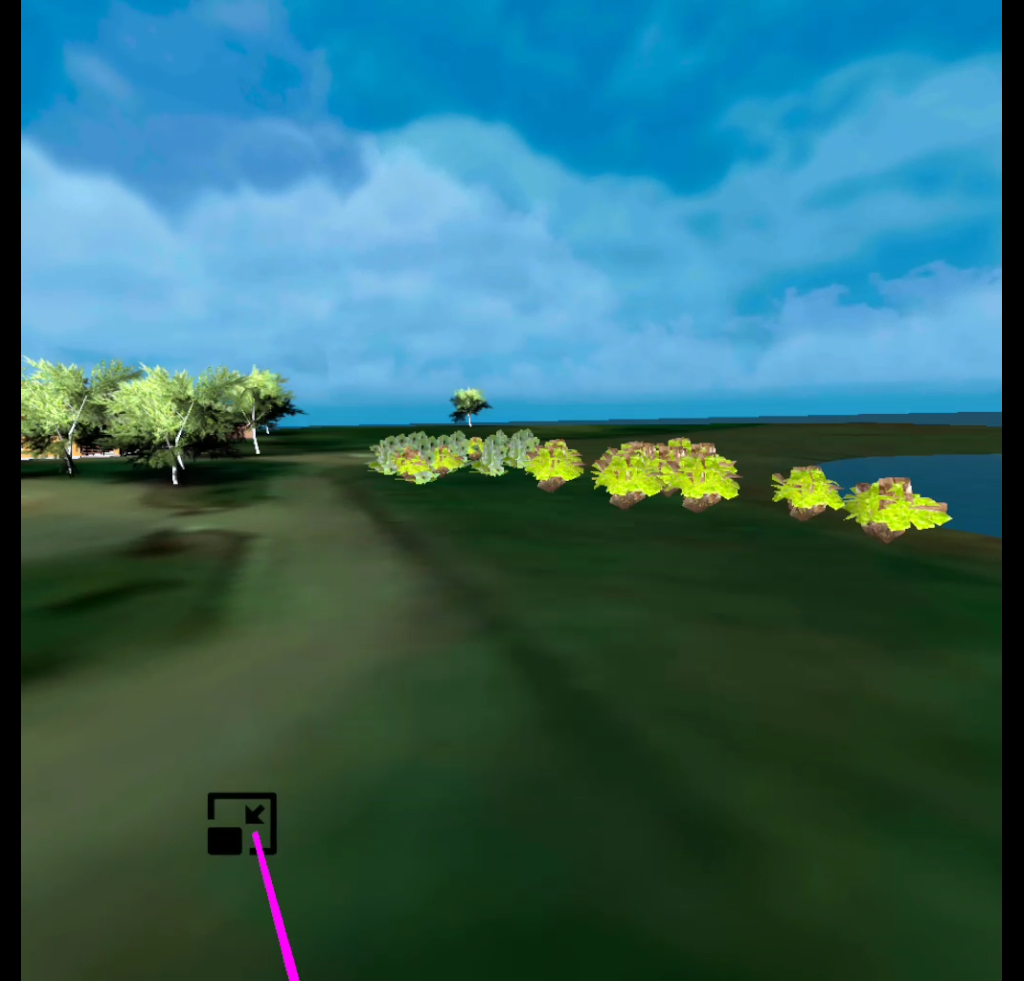}
}

\caption{(a) operation instructions, (b) menu map for choosing point of interests, (c) the sea level in example position at 2021, slider is interacted with the laser from controller, (d) sea level changed till 2100, achieved by dragging the slider button through laser pointing, (e) UI sub information main panel for relevant environmental information selection, (f) expanded text information panel example, (g) user interface before clicking 'hide UI' button, (f) user interface after clicking 'hide UI' button.}
\label{fig:fig2}
\end{figure}

\subsection{Scenes and Ocean Simulation with Geography Information System}

For the sea level simulation of each point of interest, before generated the ocean model, this project combined the prediction data files with local Digital Elevation Model (DEM) in QGIS (an Open-source cross-platform desktop geographic information system), based on overlapping regions and geographical location coordinates to determine the results of sea level rise simulation (Fig.5). 

Afterwards, on the generated terrain model with DEM, the texture from satellite imagery is added to help constructing the residential objects (Fig.6). Then with reference to local photos and the specific locations marked from satellite imagery texture, the buildings and plants of the community were modeled and put on the terrain model to constitute the land models part. After land models were finished, the ocean models generated with animated sea surface visual effect can dynamically be changed to the position specific year from 2021 to 2100 generated according to the relative positions obtained from the geographic information system (Fig.7).

\begin{figure}[htb]
\centering 
\includegraphics[width=0.49\textwidth]{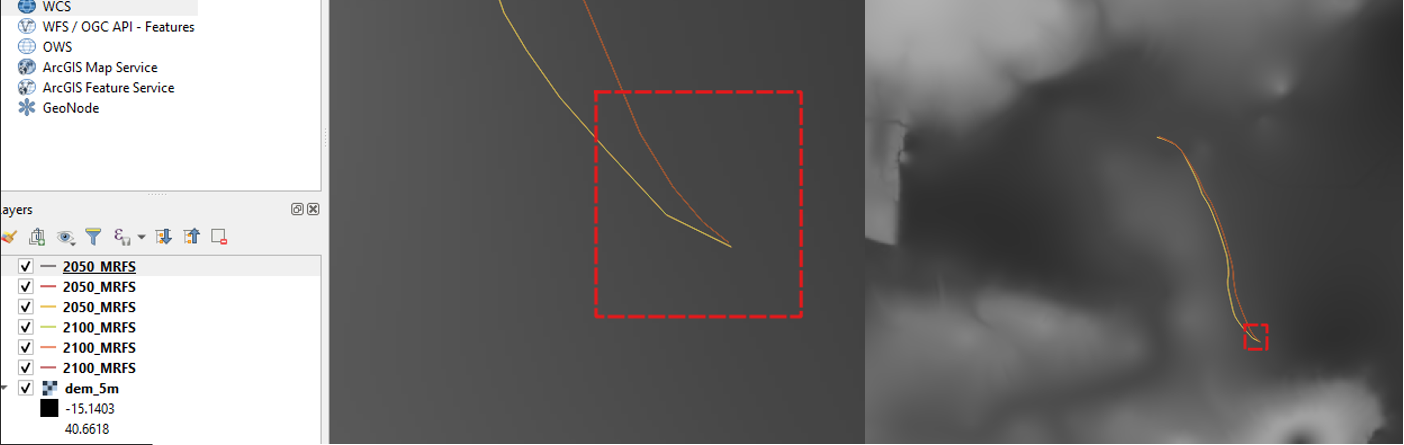} 
\caption{The DEM and prediction data overlapping.} 
\label{fig:fig3} 
\end{figure}

\begin{figure}[htb]
\centering 
\includegraphics[width=0.49\textwidth]{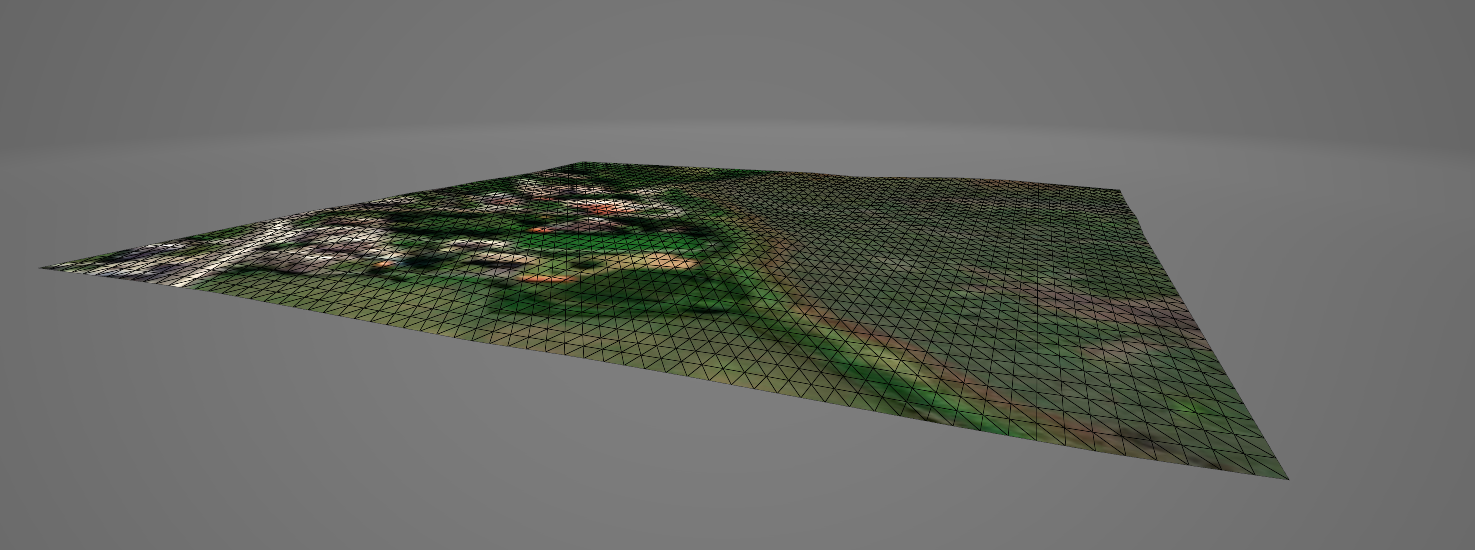} 
\caption{Generated terrain model.} 
\label{fig:fig4} 
\end{figure}

\begin{figure}[htb]
\centering 
\includegraphics[width=0.49\textwidth]{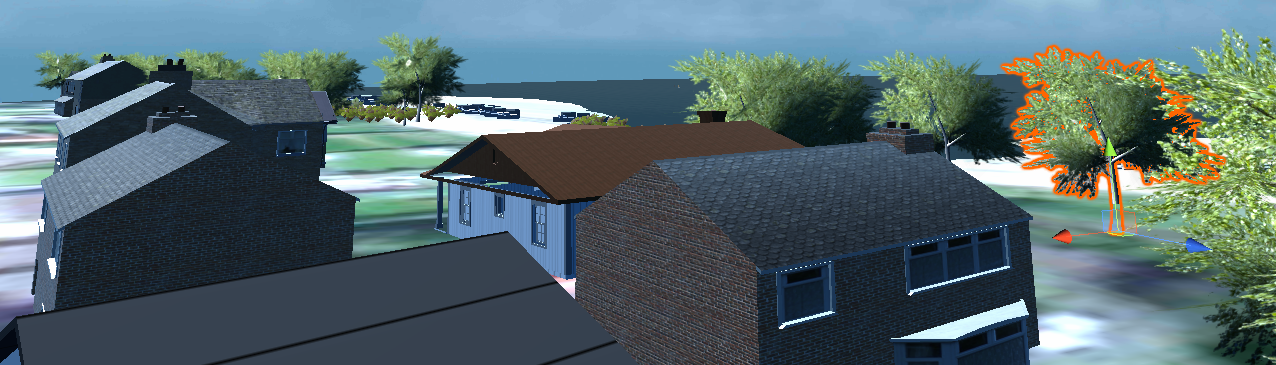} 
\caption{Built residential scene in Unity3D.} 
\label{fig:fig5} 
\end{figure}

Finally, to achieve a fully integrated scene, additional  relevant geographical and ecological information was added to the sub-panel, along with sound simulation of the changes in the sound of waves rising and falling with the sea level, which further enhanced the user's sense of immersion. The importance of sound in human-computer-interaction is often ignored but critical when you are fully immersing a user within a virtual space \cite{rogers2018vanishing}. 

\section{Experiments and Future Work}
Because of the Covid-19, only one initial feedback session was done with our council partners which included around ten people under strict COVID restrictions. The feedback season was in Fingal county council building in  Swords, Dublin, Ireland and no participants tested positive for COVID subsequently, this does not definite prove that our safety regime was successful but helped advise us in our preparation for an experiment  with both local residence and members of the public in the future. The feedback session used a CX 1 CleanBox which uses Ultra violet(UV) light to clean a Head Mounted Display(HMD). The HMD's used in this experiment where also pretreated with an Superhydrophobic nanoparticle solution that allowed them to be cleaned just using UV between users. All full experiment is planned in Q2 of 2022. This experiment will be conducted with local residents in Portrane to test whether they can have a clearer understanding of the climate change education and how their attitude towards climate change would change assisted by virtual reality, with the control group of individuals who will not be directly effected.

\subsection{Abbreviations and Acronyms}\label{AA}
Define abbreviations and acronyms the first time they are used in the text, 
even after they have been defined in the abstract. Abbreviations such as 
IEEE, SI, MKS, CGS, ac, dc, and rms do not have to be defined. Do not use 
abbreviations in the title or heads unless they are unavoidable.

\begin{figure}[htb]
\centering 
\subfigure[ ]{
\includegraphics[width=3.75cm]{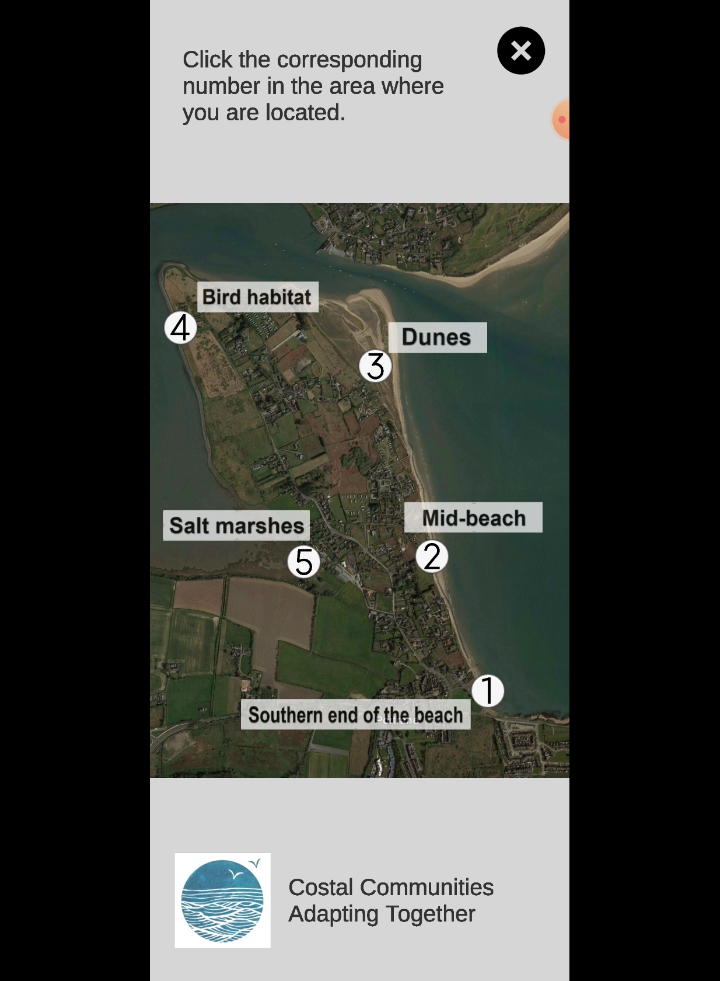}
%\caption{fig1}
}
\quad
\subfigure[ ]{
\includegraphics[width=3.75cm]{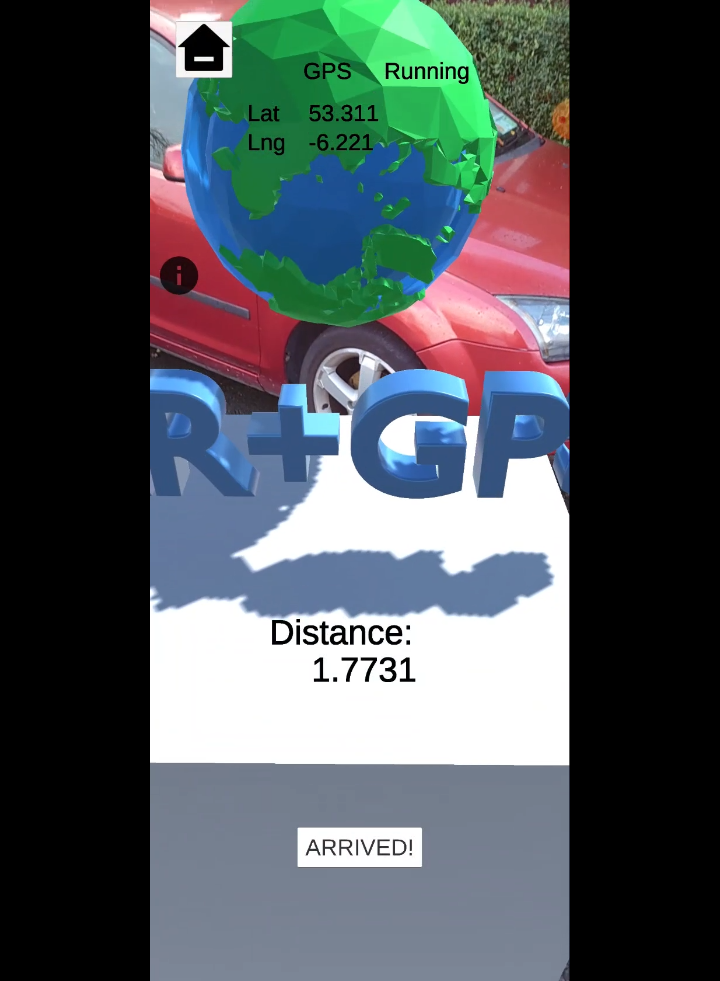}
}
\quad
\subfigure[ ]{
\includegraphics[width=3.75cm]{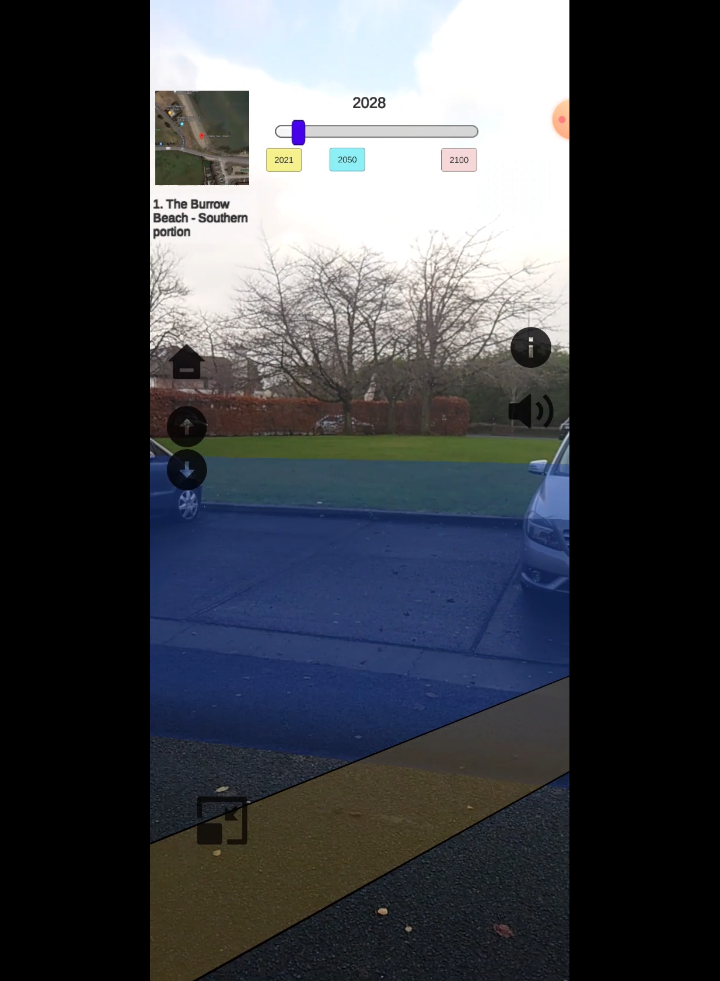}
} 
\quad
\subfigure[ ]{
\includegraphics[width=3.75cm]{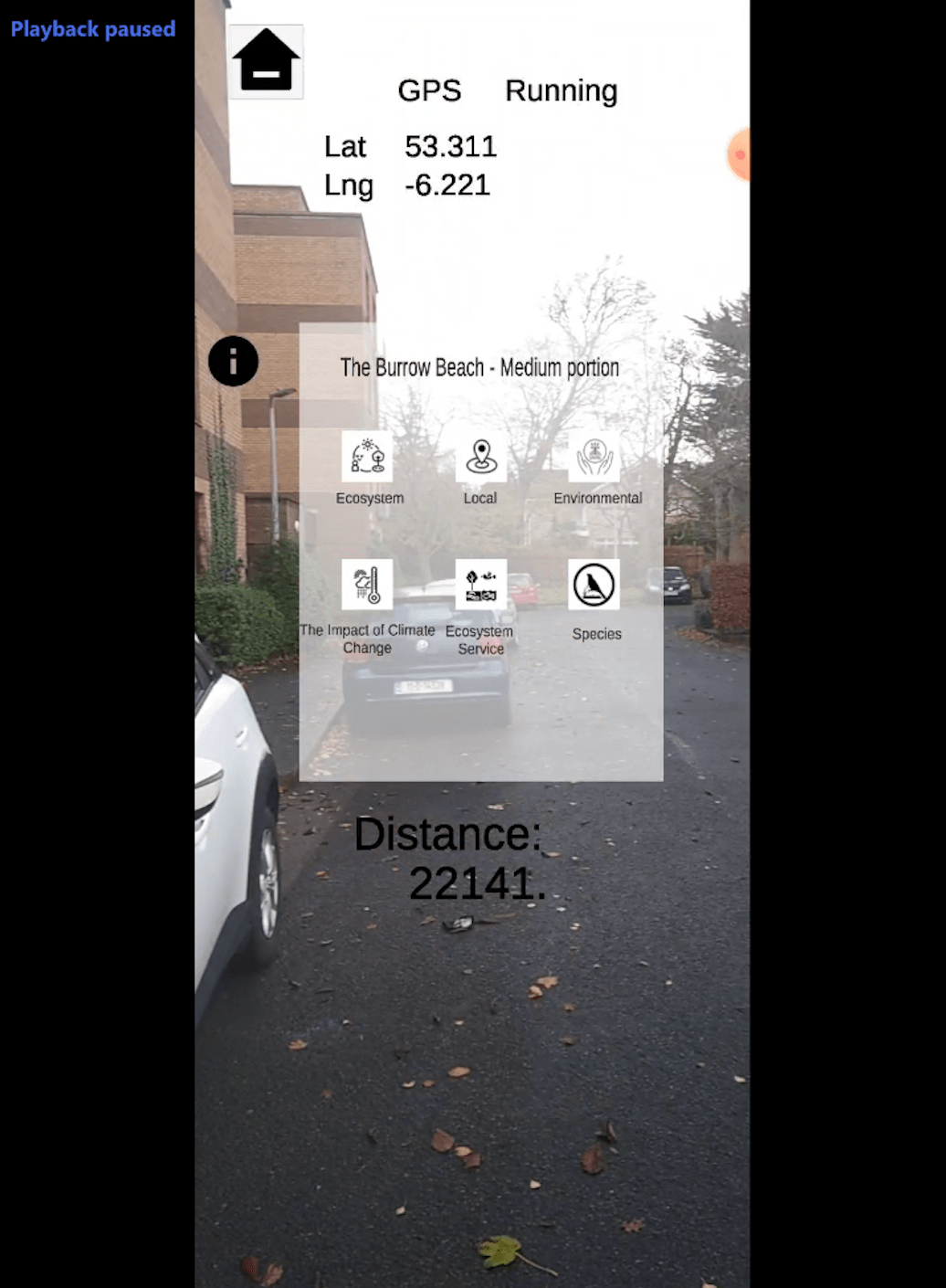}
}
\caption{AR mobile application demo. (a) menu page of the app, (b) navigation function that can lead the user to the preset target position of selected point of interests, (c) ocean simulation function that generates the ocean model to cover the real sea surface to show the sea level rise prediction, (d) inherited from the VR version of the information display function.} 
\label{fig:fig6} 
\end{figure}

Meanwhile, based on the user interface and the predicted geographic location information of ocean, an Augmented Reality mobile application is under developing now (Fig.8), it is also developed with Unity3D game engine and uses ARFoundation and Global Position System (GPS) to realize the plane detect and simulated ocean model placement in actual location of Portrane. The user interfaces and application structure logic are migrated from VR project in order to ensure the function consistency of the projects. This application will be tested in 2022 and further experiments will be conducted to allow for a comparison between the VR and AR versions of this application. 

The experimental survey content is being designed in order to learn the advantages and disadvantages of such platform transplantation from user feedback, as well as what specific adjustments should be made to the user interface design at the human-computer interaction level to adapt to the operation scenarios of AR and VR respectively. Xuanhui Xu \textit{et al.} ~\cite{inproceedings} (Fig.9) designed a experiment questionnaire based on a project of achieving the conversion of a veterinary teaching workstation VR tool to mobile devices and initially explores the performance, this is a very good demonstration and can be used to conduct experimental investigations based on this design.

\begin{figure}[htb]
\centering 
\includegraphics[width=0.49\textwidth]{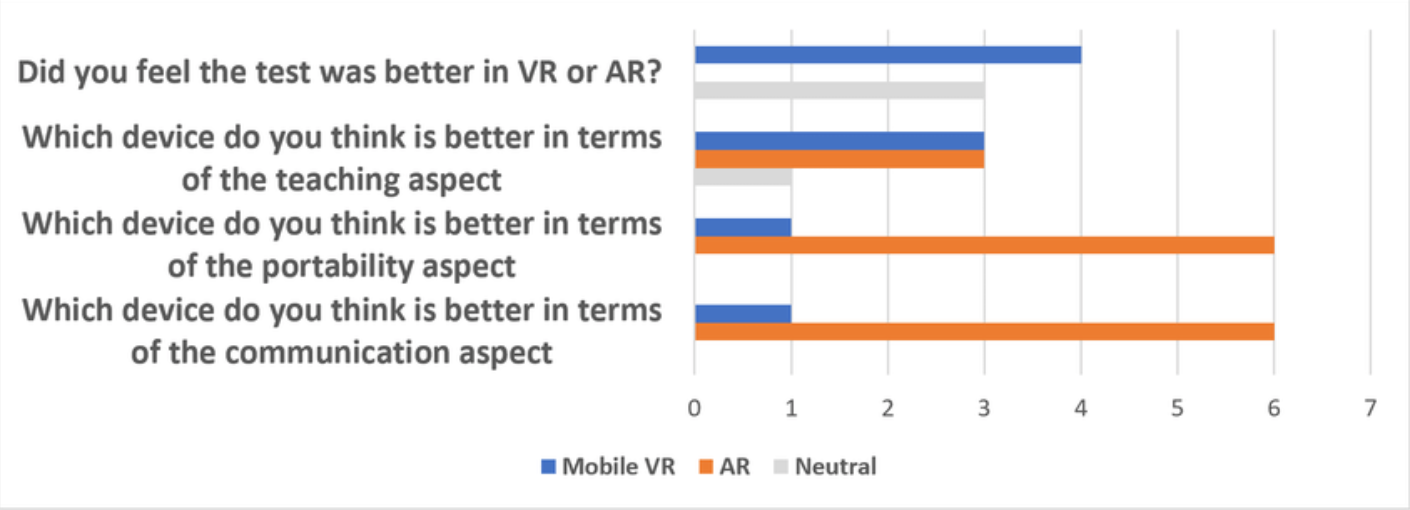} 
\caption{Experiment questionnaire design from Xuanhui Xu \textit{et al.}~\cite{inproceedings}.} 
\label{fig:fig9} 
\end{figure}

Uses this approach based on quantitative analysis of the feedback data, an evaluation will be then conducted to assess application applicability to help more intuitively bring the urgency of the climate change crisis to people who will be effect in the next 50 years.

\section{Conclusion}

Education is critical to raise awareness on climate change \cite{unesco2020education}, and to help the public understand sometimes quite complex concepts within Environmental Sciences, visualization is the key. This work has illustrated how an environmental AR/VR educational application can be created and how an initial feedback session was conducted during strict COVID restrictions. 

Currently the next step for this project is to evaluate an AR version of the application and subsequently explore if VR and/or AR is a superior medium than other mediums for climate education. 

\section*{Acknowledgment}

This research forms part of the  CONSUS Programme which is funded under the SFI   trategic Partnerships Programme   (16/SPP/3296) and is co-funded by Origin Enterprises Plc. 

% \section*{References}

% Please number citations consecutively within brackets \cite{b1}. The 
% sentence punctuation follows the bracket \cite{b2}. Refer simply to the reference 
% number, as in \cite{b3}---do not use ``Ref. \cite{b3}'' or ``reference \cite{b3}'' except at 
% the beginning of a sentence: ``Reference \cite{b3} was the first $\ldots$''

% Number footnotes separately in superscripts. Place the actual footnote at 
% the bottom of the column in which it was cited. Do not put footnotes in the 
% abstract or reference list. Use letters for table footnotes.

% Unless there are six authors or more give all authors' names; do not use 
% ``et al.''. Papers that have not been published, even if they have been 
% submitted for publication, should be cited as ``unpublished'' \cite{b4}. Papers 
% that have been accepted for publication should be cited as ``in press'' \cite{b5}. 
% Capitalize only the first word in a paper title, except for proper nouns and 
% element symbols.

% For papers published in translation journals, please give the English 
% citation first, followed by the original foreign-language citation \cite{b6}.

\balance


% \bibitem{b1} G. Eason, B. Noble, and I. N. Sneddon, ``On certain integrals of Lipschitz-Hankel type involving products of Bessel functions,'' Phil. Trans. Roy. Soc. London, vol. A247, pp. 529--551, April 1955.
% \bibitem{b2} J. Clerk Maxwell, A Treatise on Electricity and Magnetism, 3rd ed., vol. 2. Oxford: Clarendon, 1892, pp.68--73.
% \bibitem{b3} I. S. Jacobs and C. P. Bean, ``Fine particles, thin films and exchange anisotropy,'' in Magnetism, vol. III, G. T. Rado and H. Suhl, Eds. New York: Academic, 1963, pp. 271--350.
% \bibitem{b4} K. Elissa, ``Title of paper if known,'' unpublished.
% \bibitem{b5} R. Nicole, ``Title of paper with only first word capitalized,'' J. Name Stand. Abbrev., in press.
% \bibitem{b6} Y. Yorozu, M. Hirano, K. Oka, and Y. Tagawa, ``Electron spectroscopy studies on magneto-optical media and plastic substrate interface,'' IEEE Transl. J. Magn. Japan, vol. 2, pp. 740--741, August 1987 [Digests 9th Annual Conf. Magnetics Japan, p. 301, 1982].
% \bibitem{b7} M. Young, The Technical Writer's Handbook. Mill Valley, CA: University Science, 1989.
% \end{thebibliography}
\vspace{12pt}
\color{red}
% IEEE conference templates contain guidance text for composing and formatting conference papers. Please ensure that all template text is removed from your conference paper prior to submission to the conference. Failure to remove the template text from your paper may result in your paper not being published.

\end{document}